\documentclass[pra,preprint,eqsecnum,showpacs]{revtex4}
\usepackage{amsmath}
\usepackage{graphicx}
\usepackage{bm}
\begin{document}
\title{Coherent states for the quantum mechanics on a torus}
\author{K. Kowalski and J. Rembieli\'nski}
\affiliation{Department of Theoretical Physics, University
of \L\'od\'z, ul.\ Pomorska 149/153, 90-236 \L\'od\'z,
Poland}
\begin{abstract}
The coherent states for the quantum mechanics on a torus and their
basic properties are discussed.
\end{abstract}
\maketitle
\section{Introduction}
Coherent states have attracted much attention in different branches
of physics \cite{1}.  In spite of their importance the theory of coherent
states in the case when the configuration space has nontrivial
topology is far from complete.  For example the coherent states for
a quantum particle on a circle \cite{2,3} and the sphere \cite{4} have been
introduced very recently.  For an excellent review of quantum mechanics
on a circle including the coherent states we refer to very recent paper \cite{5}.  
We remark that is no general method for construction of coherent states for a
particle on an arbitrary manifold.  As a matter of fact, there
exists a general algorithm introduced by Perelomov \cite{6} of construction of 
coherent states for homogeneous spaces $X$ which are quotiens: $X=G/H$ 
of a Lie-group manifold $G$ and the stability subgroup $H$.
Unfortunately, in many cases interesting from the physical point of
view such as a particle on a circle, sphere or torus, the phase space whose 
points should label the coherent states,  more precisely a cotangent bundle
$T^*M$, where $M$ is the configuration space, is not a homogeneous space.  
In view of the lack of the general method for the construction of coherent states 
one is forced to study each case of a particle on a concrete manifold separately.
As far as we are aware the most general approach 
involving the case of the $n$-dimensional sphere $S^n$ as a configuration space
was recently introduced by Hall \cite{7}.
In this work we study the coherent states for the quantum mechanics on a torus 
based on the construction of coherent states for a particle on a circle as a 
solution of some eigenvalue equation \cite{2}.  As a matter of fact, some 
preliminary results concerning coherent states for the torus utilizing the Zak
transform have been described in section 5 of ref. 3.  Nevertheless,
the exposition of the subject presented herein is much more complete.
In section 2 we recall the construction of the coherent states for a
quantum particle on a circle \cite{2}. Section 3 deals with the quantum
mechanics on a torus.  Section 4 is devoted to the definition of the
coherent states for the torus and discussion of their basic
properties.
\section{Preliminaries --- coherent states for a quantum particle on a circle}
In this section we collect the basic facts about the coherent
states for the quantum mechanics on a circle.  These states are
related to the algebra of the form
\begin{equation}
%<2.1>
[J,U] = U,\qquad [J,U^\dagger]=-U^\dagger ,
\end{equation}
where $J$ is the angular momentum operator, $U=e^{i\hat\varphi}$ 
is the unitary operator representing the position of a quantum 
particle on a (unit) circle and we set $\hbar=1$.  Consider the 
eigenvalue equation
\begin{equation}
%<2.2>
J|j\rangle = j|j\rangle.
\end{equation}
The operators $U$ and $U^\dagger $ act on vectors $|j\rangle$ as the
ladder operators, namely
\begin{equation}
%<2.3>
U|j\rangle = |j+1\rangle,\qquad U^\dagger |j\rangle = |j-1\rangle.
\end{equation}

Demanding the time-reversal invariance of the algebra (2.1) we find
that the eigenvalues $j$ of the operator $J$ can be only integer and 
half-integer \cite{2}.

The coherent states for the quantum mechanics on a circle can be
defined by means of the eigenvalue equation \cite{2}
\begin{equation}
%<2.4>
Z|z\rangle = z|z\rangle,
\end{equation}
where 
\begin{equation}
%<2.5>
Z = e^{-J +\case12}U
\end{equation}
and the complex number
\begin{equation}
%<2.6>
z = e^{-l + i\alpha}
\end{equation}
parametrizes the circular cylinder $S^1\times{\bm R}$ which is the
classical phase space for a particle moving in a circle.  The coherent 
states specified by (2.4) can be alternatively obtained by means of
the Zak transform \cite{3}.  The projection of the vectors $|z\rangle$
onto the basis vectors $|j\rangle$ is given by
\begin{equation}
%<2.7>
\langle j|z\rangle = z^{-j}e^{-\frac{j^2}{2}}.
\end{equation}
Using the parameters $l$, and $\varphi$ we can write (2.7) as
\begin{equation}
%<2.8>
\langle j|l,\alpha\rangle =
e^{lj-i j\alpha}e^{-\frac{j^2}{2}},
\end{equation}
where $|l,\alpha\rangle\equiv|z\rangle$, with $z=e^{-l + i\alpha}$.
The coherent states are not orthogonal.  We have
\begin{subequations}
%<2.9>
\begin{eqnarray}
\langle z|w\rangle &=&
\theta_3\left(\frac{i}{2\pi}\ln z^*w\bigg\vert\frac{i}
{\pi}\right),\qquad (\hbox{\rm integer case})\\
\langle z|w\rangle &=&
\theta_2\left(\frac{i}{2\pi}\ln z^*w\bigg\vert\frac{i}
{\pi}\right),\qquad (\hbox{\rm half-integer case})
\end{eqnarray}
\end{subequations}
where $\theta_3$ and $\theta_2$ are the Jacobi theta-functions
defined by
\begin{subequations}
%<2.10>
\begin{eqnarray}
\theta_3(v|\tau) &=&
\sum_{n=-\infty}^{\infty}q^{n^2}(e^{i\pi v})^{2n},\\
\theta_2(v|\tau) &=&
\sum_{n=-\infty}^{\infty}q^{(n-\frac{1}{2})^2}(e^{i\pi
v})^{2n-1},
\end{eqnarray}
\end{subequations}
where $q=e^{i\pi\tau}$ and $\hbox{Im}\,\tau>0$.

The expectation value of the angular momentum in the coherent state is
\begin{equation}
%<2.11>
\frac{\langle l,\alpha|J|l,\alpha\rangle}{\langle
l,\alpha|l,\alpha\rangle}\approx l,
\end{equation}
where the very good approximation of the relative error is [2]: 
$\Delta l/l\approx2\pi\exp(-\pi^2)\sin(2l\pi)/l$ (see also the very 
recent paper \cite{8}), so the maximal error arising in the case $l\to0$ is of 
order 0.1 per cent.  We have remarkable exact equality 
for $l$ integer or half-integer.  Therefore, the 
parameter $l$ in $z$ can be identified with the classical angular momentum.  
Furthermore, we have the following formula on the expectation value
of the unitary operator $U$ representing the position of a particle
on a circle:
\begin{equation}
%<2.12>
\frac{\langle l,\alpha|U|l,\alpha\rangle}{\langle l,\alpha|l,\alpha\rangle}
 \approx
e^{-\frac{1}{4}}e^{i\alpha}.
\end{equation}
where the approximation is very good.  On defining the relative
expectation value
\begin{equation}
%<2.13>
\langle\!\langle U\rangle\!\rangle_{(l,\alpha)} :=
\frac{\langle U\rangle_{(l,\alpha)}}{\langle U\rangle_{(0,\alpha)}},
\end{equation}
where $\langle U\rangle_{(l,\alpha)}=\langle
l,\alpha|U|l,\alpha\rangle/\langle l,\alpha|l,\alpha\rangle$, we
get
%<2.14>
\begin{equation}
\langle\!\langle U\rangle\!\rangle_{(l,\alpha)} \approx e^{i\alpha}.
\end{equation}
Therefore, the parameter $\alpha$ can be interpreted as a classical
angle.  We point out that the approximate relation (2.14) cannot
hold in the case of the expectation value $\langle U\rangle_{(l,\alpha)}$ 
because $U$ is not diagonal in the coherent state basis.
\section{Quantum mechanics on a torus}
Now, our experience with the case of the circle discussed in the
previous section, in particular the form of the algebra (2.1), and
the fact that from the topological point of view the two-torus $T^2$
can be identified with the product of two circles, indicates the
following algebra adequate for the study of the motion on a torus:
\begin{eqnarray}
%<3.1>
&&[J_i,U_j]=\delta_{ij}U_j,\qquad
[J_i,U_j^\dagger]=-\delta_{ij}U_j^\dagger ,\\\nonumber
&&[J_i,J_j]=[U_i,U_j]=[U_i^\dagger,U_j^\dagger]=[U_i,U_j^\dagger]=0,
\qquad i,\,j=1,\,2.
\end{eqnarray}
We point out that a version of the algebra (3.1) satisfied by
$J_i$'s and the cosine and sine of the angle operators such that
\begin{equation}
%<3.2>
\cos\hat\varphi_i=\hbox{$\scriptstyle 1\over
2$}(U_i+U_i^\dagger),\qquad \sin\hat\varphi_i=\hbox{$\scriptstyle 1\over
2i$}(U_i-U_i^\dagger)
\end{equation}
were originally introduced in the context of the quantum mechanics
on a torus by Isham \cite{9} (see also \cite{10}).
Consider the eigenvalue equations
\begin{equation}
%<3.3>
{\bm J} |{\bm j}\rangle = {\bm j}|{\bm j}\rangle,
\end{equation}
where ${\bm J}=(J_1,J_2)$, and ${\bm j}=(j_1,j_2)$.  From (3.1) and
(3.3) it follows that the operators $U_i$ and $U_j^\dagger$,
$i,\,j~=~1,\,2$, act on the vectors $|{\bm j}\rangle$ as the ladder
operators, i.e.\ we have
\begin{equation}
%<3.4>
U_i|{\bm j}\rangle = |{\bm j}+{\bm e}_i\rangle,\qquad 
U_i^\dagger|{\bm j}\rangle = |{\bm j}-{\bm e}_i\rangle,
\end{equation}
where ${\bm e}_1=(1,0)$, and ${\bm e}_2=(0,1)$ are the unit vectors.  By (3.4)
we can generate the whole basis $\{|{\bm j}\rangle\}$ of the Hilbert
space of states from the unique vector $|{\bm j}_0\rangle$, where
$j_{0i}\in[0,1)$, $i=1,\,2$.  Evidently, representations with
different values of ${\bm j}_0$ are nonequivalent.  Now, let $T$ be
the anti-unitary operator of time inversion.  We have
\begin{equation}
%<3.5>
TJ_iT^{-1}=-J_i,\qquad TU_iT^{-1}=U_i^\dagger,\qquad TU_i^\dagger
T^{-1}=U_i 
\end{equation}
implying the invariance of the algebra (3.1) under time inversion.
Further, relations (3.3) and (3.4) imply
\begin{equation}
%<3.6>
T|{\bm j}\rangle= |-{\bm j}\rangle.
\end{equation}
From (3.6) it follows that $T$ is well defined on the Hilbert space
of states spanned by the vectors $|{\bm j}\rangle$ if and only if
the spectrum of ${\bm J}$ is symmetric with respect to ${\bf
0}~=~(0,0)$.  In view of (3.6) this means that $j_{01}=0$ or
$j_{01}=\case12$ and $j_{02}=0$ or $j_{02}=
\case12$.  Clearly $j_{0i}=0$
($j_{0i}=\case12$) implies integer (half-integer) eigenvalues $j_i$.
Thus, it turns out that demanding the time-reversal invariance we
have four possibilities left: $j_1$-integer and $j_2$-integer,
$j_1$-integer and $j_2$-half-integer, $j_1$-half-integer and
$j_2$-integer, and $j_1$-half-integer and $j_2$-half-integer.  These
cases will be symbolically designated by (0,0), (0,$\case12$), 
($\case12$,0) and ($\case12$,$\case12$), 
respectively, throughout this work.  We point out that in
opposition to the quantum mechanics on a circle \cite{2} the physical
interpretation of the spectrum of the angular momentum operator
${\bm J}$ is not obvious.  For example both cases (0,0) and 
$(\case12,\case12)$ seem to correspond to integer spin of a particle, 
however it is not clear what is the physical difference between them.
We finally write down the orthogonality and completeness conditions 
satisfied by the vectors $|{\bm j}\rangle$ of the form
\begin{eqnarray}
%<3.7>
&&\langle {\bm j}|{\bm j}'\rangle = \delta_{j_1j'_1}\delta_{j_2j'_2},\\
&&\sum\limits_{{\bm j}\in{\bm Z}^2} |{\bm j}\rangle\langle {\bm j}|=I,
\end{eqnarray}
where ${\bm Z}$ is the set of integers and the substitution $j_2\to
j_2-\case12$, $j_1\to j_1-\case12$, and $j_1\to j_1-\case12$ and $j_2\to
j_2-\case12$ in the cases 
$(0,\case12)$, $(\case12,0)$ and $(\case12,\case12)$, respectively, is understood.

We now discuss the coordinate representation for the quantum
mechanics on a torus.  Consider the common eigenvectors $
|\bm{\varphi}\rangle$ of the operators $U_k$ representing the position
of a particle on a torus such that 
\begin{equation}
%<3.9>
U_k|\bm{\varphi}\rangle = e^{i\varphi_k}|{\bm{\varphi}}\rangle,\qquad
k=1,\,2.
\end{equation}
These vectors form the complete set.  The resolution of the identity
can be written as
\begin{equation}
%<3.10>
\frac{1}{4\pi^2}\int\limits_{0}^{2\pi}d\varphi_1\int\limits_{0}^{2\pi}
d\varphi_2 |\bm{\varphi}\rangle\langle\bm{\varphi}|=I.
\end{equation}
If we treat torus $T^2$ as a product of two circles, that is we
restrict to the topological aspects of the torus, then completeness
gives rise to a functional representation of vectors of the form
\begin{equation}
%<3.11>
\langle f|g\rangle = \frac{1}{4\pi^2}\int\limits_{0}^{2\pi}d\varphi_1
\int\limits_{0}^{2\pi}d\varphi_2f^*(\bm{\varphi})g(\bm{\varphi}),
\end{equation}
where $f(\bm{\varphi})=\langle \bm{\varphi}|f\rangle$.  Since the basis
vectors $ |{\bm j}\rangle$ are represented by the functions
\begin{equation}
%<3.12>
e_{\bm j}(\bm{\varphi})=\langle \bm{\varphi}|\bm j\rangle = e^{i{\bm
j}\mbox{\boldmath$\scriptstyle{\cdot}$}\bm{\varphi}},
\end{equation}
where ${\bm u}\mbox{\boldmath${\cdot}$}{\bm v}=\sum_{i=1}^{2}u_iv_i$, following directly
from (3.9) and (3.4), therefore the functions which are the elements
of the Hilbert space specified by the scalar product (3.11) are
accordingly to (3.12) periodic or antiperiodic ones in $\varphi_1$ and
$\varphi_2$.  The operators $J_i$ and $U_j$, $i,\,j=1,\,2$ act in the 
representation (3.11) in the following way:
\begin{equation}
%<3.13>
J_kf(\bm{\varphi})=-i \frac{\partial f}{\partial\varphi_k},\qquad
U_kf(\bm{\varphi})=e^{i\varphi_k}f(\bm{\varphi}),\qquad k=1,\,2.
\end{equation}
We now return to (3.11).  An alternative functional representation
arises when the torus is viewed as a two-dimensional surface
embedded in ${\bm R}^3$ defined by
\begin{eqnarray}
%<3.14>
x_1 &=& (R+r\cos\varphi_2)\cos\varphi_1,\\\nonumber
x_2 &=& (R+r\cos\varphi_2)\sin\varphi_1,\\
x_3 &=& r\sin\varphi_2,\nonumber
\end{eqnarray}
where $\varphi_1$ is the azimuthal angle and $\varphi_2$ polar
angle; $R$ and $r$ are the outer and inner radius of the torus,
respectively.  In such a case the scalar product is given by \cite{11}
\begin{equation}
%<3.15>
\langle\tilde f|\tilde g\rangle = \frac{1}{4\pi^2}\int\limits_{0}^{2\pi}
d\varphi_1\int\limits_{0}^{2\pi}d\varphi_2(1+(r/R)\cos\varphi_2)
{\tilde f}^*(\bm{\varphi})\tilde g(\bm{\varphi}),
\end{equation}
where the measure $(1+(r/R)\cos\varphi_2)d\varphi_1d\varphi_2$
coincides up to the multiplicative normalization constant with the
surface element of the torus
$dS=r(R+r\cos\varphi_2)d\varphi_1d\varphi_2$. The representations
(3.11) and (3.15) are isomorphic.  The unitary operator mapping
(3.11) into (3.15) is of the form
\begin{equation}
%<3.16>
Vf(\varphi_1,\varphi_2)=\tilde f(\varphi_1,\varphi_2)=
\frac{f(\varphi_1,\varphi_2)}{\sqrt{1+(r/R)\cos\varphi_2}}.
\end{equation}
Using (3.16) and (3.13) we find that the operators act in the
representation (3.15) as follows
\begin{eqnarray}
%<3.17>
&&\tilde J_1\tilde f(\bm{\varphi})=VJ_1V^{-1}\tilde f(\bm{\varphi})=
J_1\tilde f(\bm{\varphi})=-i\frac{\partial\tilde
f}{\partial\varphi_1},\\
&&\tilde J_2\tilde f(\bm{\varphi})=VJ_2V^{-1}\tilde f(\bm{\varphi})=
-i\frac{\partial\tilde f}{\partial\varphi_2}+\frac{i}{2}
\frac{(r/R)\sin\varphi_2}{1+(r/R)\cos\varphi_2}\tilde f,\\
&&\tilde U_k\tilde f(\bm{\varphi})=VU_kV^{-1}\tilde f(\bm{\varphi})=U_k\tilde f(\bm{\varphi})=
e^{i\varphi_k}\tilde f,\qquad k=1,\,2.
\end{eqnarray}
We finally point out that the probability density for the
coordinates in the normalized state $|f\rangle$ does not depend on
the choice of the representation (3.11) or (3.15).  Indeed, we have
\begin{eqnarray}
%<3.20>
\frac{1}{4\pi^2}\int\limits_{\varphi_1}^{\varphi_1+\Delta\varphi_1}d\varphi_1
\int\limits_{\varphi_2}^{\varphi_2+\Delta\varphi_2}d\varphi_2
(1+(r/R)\cos\varphi_2)|\tilde f(\bm{\varphi})|^2\\\nonumber
{}=\frac{1}{4\pi^2}\int\limits_{\varphi_1}^{\varphi_1+\Delta\varphi_1}d\varphi_1
\int\limits_{\varphi_2}^{\varphi_2+\Delta\varphi_2}d\varphi_2|f(\bm{\varphi})|^2=
|f(\bm{\varphi})|^2\Delta\varphi_1\Delta\varphi_2,
\end{eqnarray}
where $\Delta\varphi_1\ll1$ and $\Delta\varphi_2\ll1$.
\section{Coherent states for the quantum mechanics on a torus}
\subsection{Definition of coherent states}
Based on a form of (2.5) we define the coherent states for the
quantum mechanics on a torus as the solution of the eigenvalue
equation such that
\begin{equation}
%<4.1>
{\bm Z} |{\bm z}\rangle = {\bm z}|{\bm z}\rangle,
\end{equation}
where ${\bm z}=(z_1,z_2)\in{\bm C}^2$, and
\begin{equation}
%<4.2>
Z_i=e^{-J_i+\case12}U_i,\qquad i=1,\,2.
\end{equation}
Taking into account (4.1), (3.3) and (3.4) we get
\begin{equation}
%<4.3>
\langle {\bm j}|{\bm z}\rangle = z_1^{-j_1}z_2^{-j_2}e^{-\case12{\bm j}^2}.
\end{equation}
Therefore, the coherent state $|{\bm z}\rangle$ is given by
\begin{equation}
%<4.4>
|{\bm z}\rangle = \sum\limits_{{\bm j}\in{\bm Z}^2}z_1^{-j_1}z_2^{-j_2}
e^{-\case12{\bm j}^2}|{\bm j}\rangle.
\end{equation}
Using (4.3) and (3.8) we find that the overlap of the coherent
states is
\begin{subequations}
\begin{eqnarray}
%<4.5>
\langle {\bm z}|{\bm w}\rangle &=&
\theta_3\left(\frac{i}{2\pi}\ln z_1^*w_1\bigg\vert\frac{i}{\pi}\right)
\theta_3\left(\frac{i}{2\pi}\ln z_2^*w_2\bigg\vert\frac{i}{\pi}\right)
,\qquad (\hbox{\rm (0,0) case})\\
\langle {\bm z}|{\bm w}\rangle &=&
\theta_3\left(\frac{i}{2\pi}\ln z_1^*w_1\bigg\vert\frac{i}{\pi}\right)
\theta_2\left(\frac{i}{2\pi}\ln z_2^*w_2\bigg\vert\frac{i}{\pi}\right)
,\qquad (\hbox{\rm (0,$\case12$) case})\\
\langle {\bm z}|{\bm w}\rangle &=&
\theta_2\left(\frac{i}{2\pi}\ln z_1^*w_1\bigg\vert\frac{i}{\pi}\right)
\theta_3\left(\frac{i}{2\pi}\ln z_2^*w_2\bigg\vert\frac{i}{\pi}\right)
,\qquad (\hbox{\rm ($\case12$,0) case})\\
\langle {\bm z}|{\bm w}\rangle &=&
\theta_2\left(\frac{i}{2\pi}\ln z_1^*w_1\bigg\vert\frac{i}{\pi}\right)
\theta_2\left(\frac{i}{2\pi}\ln z_2^*w_2\bigg\vert\frac{i}{\pi}\right)
.\qquad (\hbox{\rm ($\case12$,$\case12$) case})
\end{eqnarray}
\end{subequations}
Now, the phase space for a quantum particle on a torus is the
cotangent bundle $T^*T^2=T^2\times{\bm R}^2$ which is topologically
equivalent to the product of two cylinders $(S^1\times{\bm R})\times
(S^1\times{\bm R})$.  Therefore, we can use the parametrization
(2.6) and write the coordinates of the vector ${\bm z}$ labelling
the phase space for a quantum particle on a torus as
\begin{equation}
%<4.6>
z_k = e^{-l_k +i\alpha_k},\qquad k=1,\,2.  
\end{equation}
The relations (4.3)--(4.5) written in the parametrization (4.6) take
the following form:
\begin{equation}
%<4.7>
\langle {\bm j}|{\bm l},\bm{\alpha}\rangle=e^{{\bm l}
\mbox{\boldmath$\scriptstyle{\cdot}$}{\bm j}-i\bm{\alpha}
\mbox{\boldmath$\scriptstyle{\cdot}$}{\bm j}}e^{-\case12{\bm j}^2},
\end{equation}
where $|{\bm l},\bm{\alpha}\rangle\equiv|{\bm z}\rangle$ with $z_k=
e^{-l_k+i\alpha_k}$, $k=1,\,2$.  Therefore we can write the
coherent state in the form
\begin{equation}
%<4.8>
|{\bm l},\bm{\alpha}\rangle=\sum\limits_{{\bm j}\in{\bm Z}^2}
e^{{\bm l}\mbox{\boldmath$\scriptstyle{\cdot}$}{\bm j}-i\bm{\alpha}
\mbox{\boldmath$\scriptstyle{\cdot}$}
{\bm j}}e^{-\case12{\bm j}^2}|{\bm j}\rangle.
\end{equation}
An immediate consequence of (4.7), (3.8) and (2.10) are the following formulae
on the scalar products (4.5) written in the parametrization (4.6):
$$\displaylines{
\langle {\bm l},\bm{\alpha}|{\bm h},\bm{\beta}\rangle =
\theta_3\left(\frac{1}{2\pi}(\alpha_1-\beta_1)-\frac{l_1+h_1}{2}\frac{i}{\pi}
\bigg\vert\frac{i}{\pi}\right)
\theta_3\left(\frac{1}{2\pi}(\alpha_2-\beta_2)-\frac{l_2+h_2}{2}\frac{i}{\pi}
\bigg\vert\frac{i}{\pi}\right),\hfill\cr
\hfill\llap{(\hbox{\rm (0,0) case})\hspace{1em}(4.9a)}\cr
\langle {\bm l},\bm{\alpha}|{\bm h},\bm{\beta}\rangle =
\theta_3\left(\frac{1}{2\pi}(\alpha_1-\beta_1)-\frac{l_1+h_1}{2}\frac{i}{\pi}
\bigg\vert\frac{i}{\pi}\right)
\theta_2\left(\frac{1}{2\pi}(\alpha_2-\beta_2)-\frac{l_2+h_2}{2}\frac{i}{\pi}
\bigg\vert\frac{i}{\pi}\right),\hfill\cr
\hfill\llap{(\hbox{\rm (0,$\case12$) case})\hspace{1em}(4.9b)}\cr
\langle {\bm l},\bm{\alpha}|{\bm h},\bm{\beta}\rangle =
\theta_2\left(\frac{1}{2\pi}(\alpha_1-\beta_1)-\frac{l_1+h_1}{2}\frac{i}{\pi}
\bigg\vert\frac{i}{\pi}\right)
\theta_3\left(\frac{1}{2\pi}(\alpha_2-\beta_2)-\frac{l_2+h_2}{2}\frac{i}{\pi}
\bigg\vert\frac{i}{\pi}\right),\hfill\cr
\hfill\llap{(\hbox{\rm ($\case12$,0) case})\hspace{1em}(4.9c)}\cr
\langle {\bm l},\bm{\alpha}|{\bm h},\bm{\beta}\rangle =
\theta_2\left(\frac{1}{2\pi}(\alpha_1-\beta_1)-\frac{l_1+h_1}{2}\frac{i}{\pi}
\bigg\vert\frac{i}{\pi}\right)
\theta_2\left(\frac{1}{2\pi}(\alpha_2-\beta_2)-\frac{l_2+h_2}{2}\frac{i}{\pi}
\bigg\vert\frac{i}{\pi}\right),\hfill\cr
\hfill\llap{(\hbox{\rm ($\case12,\case12$) case})\hspace{1em}(4.9d)}\cr}$$
\setcounter{equation}{9}%
\subsection{Coherent states and the classical phase space}
As with the coherent states $|z\rangle$ for a quantum particle on a
circle our criterion to test the correctness of the introduced coherent
states $|{\bm z}\rangle$ for the quantum mechanics on a torus will be their
closeness to the classical phase space described by the formulae
like (2.11) and (2.14).  Consider first the expectation values of
the angular momentum ${\bm J}$.  Eqs.\ (3.8), (3.3) and (4.7) taken
together yield
\begin{subequations}
\begin{eqnarray}
%<4.10>
\frac{\langle {\bm l},\bm{\alpha}|J_k|{\bm l},\bm{\alpha}\rangle}
{\langle {\bm l},\bm{\alpha}|{\bm l},\bm{\alpha}\rangle}&=&
\frac{1}{2\theta_3(\frac{i l_k}{\pi}\big\vert\frac{i}{\pi})}
\frac{d}{dl_k}\theta_3\left(\frac{i
l_k}{\pi}\bigg\vert\frac{i}{\pi}\right),\quad k=1,\,2,
\quad (\hbox{\rm (0,0) case})\\
\frac{\langle {\bm l},\bm{\alpha}|J_{1(2)}|{\bm l},\bm{\alpha}\rangle}
{\langle {\bm l},\bm{\alpha}|{\bm l},\bm{\alpha}\rangle}&=&
\frac{1}{2\theta_{3(2)}(\frac{i l_{1(2)}}{\pi}\big\vert\frac{i}{\pi})}
\frac{d}{dl_{1(2)}}\theta_{3(2)}\left(\frac{i
l_{1(2)}}{\pi}\bigg\vert\frac{i}{\pi}\right)
,\qquad (\hbox{\rm (0,$\case12$) case})\\
\frac{\langle {\bm l},\bm{\alpha}|J_{1(2)}|{\bm l},\bm{\alpha}\rangle}
{\langle {\bm l},\bm{\alpha}|{\bm l},\bm{\alpha}\rangle}&=&
\frac{1}{2\theta_{2(3)}(\frac{i l_{1(2)}}{\pi}\big\vert\frac{i}{\pi})}
\frac{d}{dl_{1(2)}}\theta_{2(3)}\left(\frac{i
l_{1(2)}}{\pi}\bigg\vert\frac{i}{\pi}\right)
,\qquad (\hbox{\rm ($\case12$,0) case})\\
\frac{\langle {\bm l},\bm{\alpha}|J_k|{\bm l},\bm{\alpha}\rangle}
{\langle {\bm l},\bm{\alpha}|{\bm l},\bm{\alpha}\rangle}&=&
\frac{1}{2\theta_2(\frac{i l_k}{\pi}\big\vert\frac{i}{\pi})}
\frac{d}{dl_k}\theta_2\left(\frac{i
l_k}{\pi}\bigg\vert\frac{i}{\pi}\right),\quad k=1,\,2.
\quad (\hbox{\rm ($\case12$,$\case12$) case})
\end{eqnarray}
\end{subequations}
Proceeding analogously as in the case of the coherent states for the
circle we find that for $l_i$-integer or half-integer
\begin{equation}
%<4.11>
\frac{\langle {\bm l},\bm{\alpha}|{\bm J}|{\bm l},\bm{\alpha}\rangle}
{\langle {\bm l},\bm{\alpha}|{\bm l},\bm{\alpha}\rangle} = {\bm l}
\end{equation}
and in general case
\begin{equation}
%<4.12>
\frac{\langle {\bm l},\bm{\alpha}|{\bm J}|{\bm l},\bm{\alpha}\rangle}
{\langle {\bm l},\bm{\alpha}|{\bm l},\bm{\alpha}\rangle} \approx {\bm l},
\end{equation}
where the approximation is very good (the maximal error is of order
0.1 per cent).  Therefore the parameter ${\bm l}$ can be regarded as
a classical angular momentum.  

We now discuss the expectation values of the unitary operators $U_i$
representing the position of a quantum particle on a torus.  On
using (3.8), (3.4) and (4.7) we arrive at the following relations
\begin{subequations}
\begin{eqnarray}
%<4.13>
\frac{\langle {\bm l},\bm{\alpha}|U_k|{\bm l},\bm{\alpha}\rangle}
{\langle {\bm l},\bm{\alpha}|{\bm l},\bm{\alpha}\rangle}&=&
e^{-\case14}e^{i \alpha_k}\frac{\theta_2(\frac{i
l_k}{\pi}\big\vert\frac{i}{\pi})}{\theta_3
(\frac{i l_k}{\pi}\big\vert\frac{i}{\pi})},\quad k=1,\,2,
\quad (\hbox{\rm (0,0) case})\\
\frac{\langle {\bm l},\bm{\alpha}|U_{1(2)}|{\bm l},\bm{\alpha}\rangle}
{\langle {\bm l},\bm{\alpha}|{\bm l},\bm{\alpha}\rangle}&=&
e^{-\case14}e^{i \alpha_{1(2)}}\frac{\theta_{2(3)}(\frac{i
l_{1(2)}}{\pi}\big\vert\frac{i}{\pi})}{\theta_{3(2)}
(\frac{i l_{1(2)}}{\pi}\big\vert\frac{i}{\pi})}
,\qquad (\hbox{\rm (0,$\case12$) case})\\
\frac{\langle {\bm l},\bm{\alpha}|U_{1(2)}|{\bm l},\bm{\alpha}\rangle}
{\langle {\bm l},\bm{\alpha}|{\bm l},\bm{\alpha}\rangle}&=&
e^{-\case14}e^{i \alpha_{1(2)}}\frac{\theta_{3(2)}(\frac{i
l_{1(2)}}{\pi}\big\vert\frac{i}{\pi})}{\theta_{2(3)}
(\frac{i l_{1(2)}}{\pi}\big\vert\frac{i}{\pi})}
,\qquad (\hbox{\rm ($\case12$,0) case})\\
\frac{\langle {\bm l},\bm{\alpha}|U_k|{\bm l},\bm{\alpha}\rangle}
{\langle {\bm l},\bm{\alpha}|{\bm l},\bm{\alpha}\rangle}&=&
e^{-\case14}e^{i \alpha_k}\frac{\theta_3(\frac{i
l_k}{\pi}\big\vert\frac{i}{\pi})}{\theta_2
(\frac{i l_k}{\pi}\big\vert\frac{i}{\pi})},\quad k=1,\,2.
\quad (\hbox{\rm ($\case12$,$\case12$) case})
\end{eqnarray}
\end{subequations}
As with the case of the coherent states for the circle it follows
that in arbitrary case
\begin{equation}
%<4.14>
\frac{\langle {\bm l},\bm{\alpha}|U_k|{\bm l},\bm{\alpha}\rangle}
{\langle {\bm l},\bm{\alpha}|{\bm l},\bm{\alpha}\rangle}\approx
e^{-\case14}e^{i \alpha_k},\qquad k=1,\,2,
\end{equation}
where the approximation is very good.  Therefore, introducing the 
relative expectation value
\begin{equation}
%<4.15>
\langle\!\langle U_k\rangle\!\rangle_{({\bm l},\bm{\alpha})} :=
\frac{\langle U_k\rangle_{({\bm l},\bm{\alpha})}}{\langle
U_k\rangle_{({\bf 0},\bm{\alpha})}},\qquad k=1,\,2,
\end{equation}
where $\langle U_k\rangle_{({\bm l},\bm{\alpha})}=
\langle {\bm l},\bm{\alpha}|U_k|{\bm l},\bm{\alpha}\rangle/
\langle {\bm l},\bm{\alpha}|{\bm l},\bm{\alpha}\rangle$, we obtain
\begin{equation}
%<4.16>
\langle\!\langle U_k\rangle\!\rangle_{({\bm l},\bm{\alpha})} 
\approx e^{i\alpha_k},\qquad k=1,\,2.
\end{equation}
Thus, it turns out that $\alpha_k$ can be regarded as a classical
angle parametrizing the position of a particle on a torus.  Another
evidence for such interpretation of the parameters $\alpha_k$ is
provided by the behaviour of the probability density for the
coordinates in the normalized coherent state.  Indeed, let us
restrict for brevity to the case (0,0).  The probability density is
then given by
\begin{equation}
%<4.17>
p_{({\bm l},\bm{\alpha})}(\bm{\varphi})=\frac{|\langle \bm{\varphi}|{\bm
l},\bm{\alpha}\rangle|^2}{\langle {\bm l},\bm{\alpha}|{\bm
l},\bm{\alpha}\rangle}=\frac{|\theta_3(\frac{1}{2\pi}(\varphi_1-\alpha_1-i l_1)
\big\vert\frac{i}{2\pi})\theta_3(\frac{1}{2\pi}(\varphi_2-\alpha_2-i l_2)
\big\vert\frac{i}{2\pi})|^2}{\theta_3(\frac{i l_1}{\pi}
\big\vert\frac{i}{\pi})\theta_3(\frac{i l_2}{\pi}
\big\vert\frac{i}{\pi})}
\end{equation}
following directly from (3.12) and (4.7).  From computer simulations
it follows that the function $p_{({\bm l},\bm{\alpha})}(\bm{\varphi})$ is
peaked at $\bm{\varphi}=\bm{\alpha}$ (see figures).  This observation
confirms once more the role of $\alpha_k$ as a classical angle.
\begin{figure*}
\centering
\includegraphics[scale=.8]{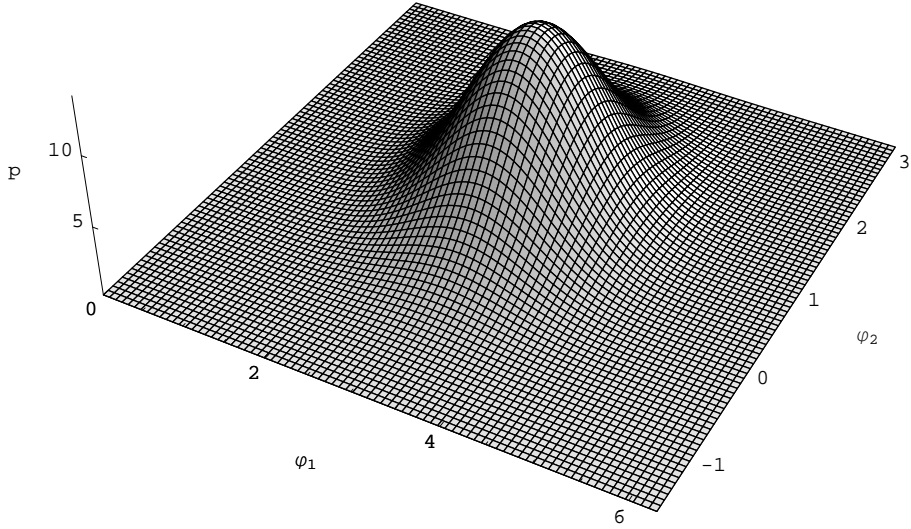}
\caption{The plot of the probability density given by (4.17) 
with $l_1=1$, $l_2=1$, $\alpha_1=\pi$ and $\alpha_2=\pi/3$.  The view
point slightly above the surface.}
\end{figure*}
\begin{figure*}
\centering
\includegraphics[scale=.8]{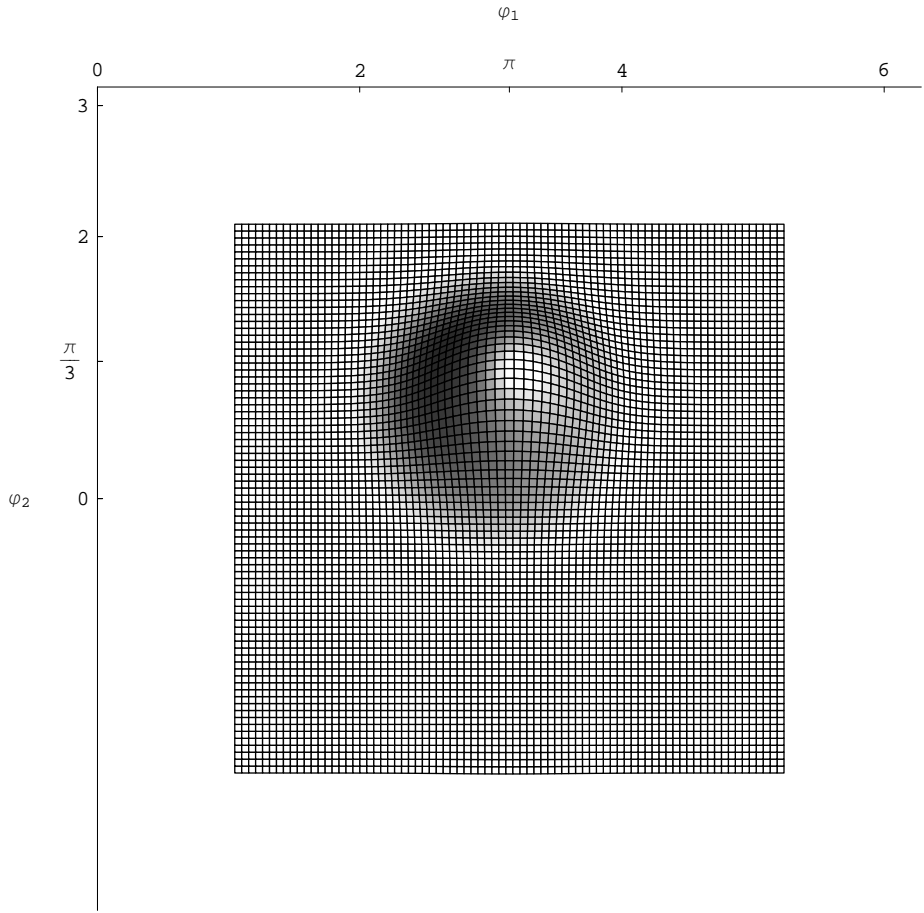}
\caption{The plot of the surface from figure 1 with the view point
directly above.  The maximum of the probability density given by
(4.17) at $\varphi_1=\pi$ and $\varphi_2=\pi/3$ is easily seen.}
\end{figure*}
\subsection{Generalization to quasiperiodic boundary conditions}
Bearing in mind the possible applications
of the introduced approach in the condensed matter physics one
should involve the case of the general ${\bm j}_0$ labelling the
ground state $|{\bm j}_0\rangle$ (see (3.4) and discussion below).
In fact, the wavefunctions which are the elements of the Hilbert
space of states specified by the scalar product (3.11) obey then the
general quasiperiodic boundary condition
\begin{equation}
%<4.18>
f(\varphi_1,\varphi_2)=e^{i2\pi j_{01}}e^{i2\pi j_{02}}
f(\varphi_1,\varphi_2),
\end{equation}
following directly from (3.12) and (3.8) with ${\bm j}$ replaced
with ${\bm j}+{\bm j}_0$.  Similar functions (Bloch functions)
appear in solid state physics.  We also point out that the quasiperiodic 
boundary condition analogous to (4.18) arises for the quantum mechanics on 
a circle.  For the general case of the quantum mechanics on multiply-connected 
spaces such condition was discussed by D\"urr {\em et al} \cite{12}.  In the context of the quantum mechanics of a free
particle on a plane with an extracted point the quasiperiodic boundary condition
was investigated by us in \cite{13}.  Clearly, the plane with an extracted point 
is also an example of a multiply-connected space.  We recall that the case of the
time-reversal symmetry discussed before implied $j_{0i}=0$ or
$j_{0i}=\case12$, $i=1,\,2$, that is the periodic and antiperiodic
functions in $\varphi_1$ and $\varphi_2$.  The generalization of the
results obtained earlier to the case of arbitrary ${\bm j}_0$ is
straightforward.  The scalar product of coherent states is given by
\begin{equation}
%<4.19>
\langle {\bm z}|{\bm w}\rangle =(z_1^*w_1)^{-j_{01}}(z_2^*w_2)^{-j_{02}}
e^{-j_{01}^2-j_{02}^2}
\theta_3\left(\frac{i}{2\pi}(\ln z_1^*w_1+2j_{01})\bigg\vert\frac{i}{\pi}\right)
\theta_3\left(\frac{i}{2\pi}(\ln z_2^*w_2+2j_{02})\bigg\vert\frac{i}{\pi}\right).
\end{equation}
The relations (4.5) can be easily obtained from (4.19).  Indeed, (4.5a) is an 
immediate consequence of (4.19).  The formulas (4.5b), (4.5c) and (4.5d) referring 
to $j_{0i}=\case12$, are implied by (4.19) and the identity
\begin{equation}
%<4.20>
\theta_{3(2)}(v+\hbox{$\scriptstyle\tau\over2$}|\tau)=
e^{-i\pi(\frac{\tau}{4}+v)}\theta_{2(3)}(v|\tau).
\end{equation}
In the parametrization (4.6) the scalar product of the coherent
states takes the form
$$\displaylines{
\langle {\bm l},\bm{\alpha}|{\bm h},\bm{\beta}\rangle =
e^{({\bm l}+{\bm h})
\mbox{\boldmath$\scriptstyle{\cdot}$}{\bm j}_0+i(\bm{\alpha}
-\bm{\beta})\mbox{\boldmath$\scriptstyle{\cdot}$}{\bm j}_0}e^{-{\bm
j}_0^2}\hfill\cr
\quad{}\times\theta_3\left(\frac{1}{2\pi}(\alpha_1-\beta_1)-\frac{l_1+h_1-2j_{01}}{2}\frac{i}{\pi}
\bigg\vert\frac{i}{\pi}\right)
\theta_3\left(\frac{1}{2\pi}(\alpha_2-\beta_2)-\frac{l_2+h_2-2j_{02}}{2}\frac{i}{\pi}
\bigg\vert\frac{i}{\pi}\right).\hfill\llap{(4.21)}\cr}$$
\setcounter{equation}{21}%
Now, proceeding as with (4.10) we arrive at the following formula on
the expectation value of the angular momentum ${\bm J}$ in the case
of an arbitrary ${\bm j}_0$:
\begin{equation}
%<4.22>
\frac{\langle {\bm l},\bm{\alpha}|J_k|{\bm l},\bm{\alpha}\rangle}
{\langle {\bm l},\bm{\alpha}|{\bm l},\bm{\alpha}\rangle}=j_{0k}+
\frac{1}{2\theta_3(\frac{i (l_k-j_{0k})}{\pi}\big\vert\frac{i}{\pi})}
\frac{d}{dl_k}\theta_3\left(\frac{i
(l_k-j_{0k})}{\pi}\bigg\vert\frac{i}{\pi}\right),\quad k=1,\,2.
\end{equation}
From (4.22) and (4.10a) it follows that the approximate relation
(4.12) holds in the case of an arbitrary ${\bm j}_0$.  It can be
easily checked that (4.10) are implied by (4.22), (4.20) and the
fact that the Jacobi theta-functions are even functions of $v$,
i.e.\ $\theta_3(-v|\tau)=\theta_3(v|\tau)$, and $\theta_2(-v|\tau)
=\theta_2(v|\tau)$.  Furthermore, using technique applied for
derivation (4.13) we get the expectation values of the unitary
operators $U_i$ representing the position of the quantum particle on
a torus such that
\begin{equation}
%<4.23>
\frac{\langle {\bm l},\bm{\alpha}|U_k|{\bm l},\bm{\alpha}\rangle}
{\langle {\bm l},\bm{\alpha}|{\bm l},\bm{\alpha}\rangle}=
e^{-\case14}e^{i \alpha_k}\frac{\theta_2(\frac{i}{\pi}
(l_k-j_{0k})\big\vert\frac{i}{\pi})}{\theta_3
(\frac{i}{\pi}(l_k-j_{0k})\big\vert\frac{i}{\pi})},\quad k=1,\,2.
\end{equation}
The formulas (4.13) can be easily derived from (4.23) with the help
of (4.20).  Taking into account (4.23) and (4.13a) we find that the
approximation relation (4.14) is valid for every ${\bm j}_0$.
Finally, one can easily obtain the following generalization of the
formula (4.17) on the probability density for the coordinates in the
normalized coherent state
\begin{equation}
%<4.24>
p_{({\bm l},\bm{\alpha})}(\bm{\varphi})=\frac{|\langle \bm{\varphi}|{\bm
l},\bm{\alpha}\rangle|^2}{\langle {\bm l},\bm{\alpha}|{\bm
l},\bm{\alpha}\rangle}=\frac{|\theta_3(\frac{1}{2\pi}(\varphi_1-
\alpha_1-i(l_1-j_{01}))
\big\vert\frac{i}{2\pi})\theta_3(\frac{1}{2\pi}(\varphi_2-
\alpha_2-i(l_2-j_{02}))
\big\vert\frac{i}{2\pi})|^2}{\theta_3(\frac{i}{\pi}(l_1-j_{01})
\big\vert\frac{i}{\pi})\theta_3(\frac{i}{\pi}(l_2-j_{02})
\big\vert\frac{i}{\pi})}.
\end{equation}
Comparing (4.24) and (4.17) we find that the probability density
(4.24) is peaked at $\bm{\varphi}=\bm{\alpha}$ for an arbitrary
${\bm j}_0$.
\section{Conclusion}
In this work we have introduced the coherent states for the quantum
mechanics on a torus.  We have not discussed in this paper the Bargmann
representation which can be easily obtained by using the relations
derived for the circle \cite{2}.  In our opinion, besides of the
possible applications in the solid state physics mentioned above, 
the observations of this work would be of importance in the theory 
of quantum chaos.  We only recall that the torus is the configuration 
space of the double pendulum and toroidal pendulum which are well 
known to show chaotic behaviour.  Another possible application of the 
constructed coherent states is nanotechnology, especially nanoscopic 
quantum rings \cite{14}.  Furthermore, we hope that similarly as with
the coherent states for a particle on a circle which have been applied
by Ashtekar {\em et al\/} \cite{15} in loop quantum gravity also introduced
coherent states for the torus would be of importance in this theory.
We finally remark that the results of this paper can be immediately 
generalized to the case of the $n$-dimensional torus.
\section*{Acknowledgements}
This paper has been supported by the Polish Ministry of Scientific 
Research and Information Technology under the grant 
No PBZ-MIN-008/P03/2003.

\end{document}